\documentclass[final,5p,times,twocolumn]{elsarticle}
\usepackage{graphicx}
\usepackage{amssymb}
\usepackage{lineno}
\usepackage{amssymb}
\usepackage[colorlinks,linkcolor=SB,citecolor=red]{hyperref}
\usepackage{geometry}
\usepackage{graphicx}
\usepackage{natbib}
\usepackage{etoolbox}
\usepackage{float}

\usepackage{hyperref}
\usepackage{float}
\usepackage[caption = false]{subfig}
\usepackage{amsmath,amssymb,amsfonts}

\apptocmd{\sloppy}{\hbadness 10000\relax}{}{}

\biboptions{comma,sort&compress}
\journal{}

\begin{document}

\begin{frontmatter}


\title{Growth Dynamics of Value and Cost Trade-off in Temporal Networks}



\author[Hasaniaddress1]{Sheida Hasani}
\author[Masoomiaddress1]{Razieh Masoomi}
\author[ardalankiaaddress1,ccnsdaddress]{Jamshid Ardalankia\corref{cor1}\fnref{label2}}
\ead{jamshid.ardalankia@gmail.com}
\author[Masoomiaddress1]{Hamid Jafari}
\author[Hasaniaddress1]{Mohammadbashir Sedighi}

\address[Hasaniaddress1]{Department of Management, Science and Technology, Amirkabir University of Technology, Tehran, Iran}
\address[Masoomiaddress1]{Department of Physics, Shahid Beheshti University, G.C., Evin, Tehran, 19839, Iran}
\address[ardalankiaaddress1]{Department of Financial Management, Shahid Beheshti University, G.C., Evin, Tehran, 19839, Iran}
\address[ccnsdaddress]{Center for Complex Networks and Social Datascience, Department of Physics, Shahid Beheshti University, G.C., Evin, Tehran, 19839, Iran}

\begin{abstract}
The question is: What does happen to the real-world networks which cause them not to grow permanently? The idea here is that real-world networks have to pay the cost of growth. We investigate the growth and trade-off between value and cost in the networks with cost and preferential attachment together. Since the preferential attachment in the BA model does not consider any stop against the infinite growth of networks, we introduce a modified version of preferential attachment of the BA model. This idea makes sense because the growth of real networks may be finite. In the present study, by combining preferential attachment in the science of temporal networks (interval graphs), and, the first-order differential equations of value and cost of making links, the future equilibrium of an evolving network is illustrated. During the process of achieving a winning position, the variables against growth such as the competition cost, besides the internally structural cost may emerge. In the end, by applying this modified model, we found the circumstances in which a trade-off between value and cost emerges.

\end{abstract}


\end{frontmatter}



\section{Introduction}
Impressive efforts respecting the concept of BA preferential attachment have been successfully carried out~\cite{Barabaacutesi1999,Barabsi1999}.
Also, it is worth mentioning that real-world networks inevitably pay the \textit{cost} of growth. That is why a certain event corresponding to growth is not out of charge. Hence, real networks may not have infinite growth. A phenomenon like friction against growth -say \textit{cost} of making new attachments- emerges in the networks with preferential attachment~\cite{Barabaacutesi1999,Newman2001,Barabsi1999,Jafarietal,shirazi2013}

Network science has a wide range of applicability in the machine learning algorithms~\cite{Kurths_ML_2020}, the social systems~\cite{shirazi2013},
the financial and economic systems~\cite{Casarin2019,Bardoscia2017,ardalankia2020mapping,NamakiBanking2020,Zhao2018,Habibnia2017}, 
systems with ordinal behaviors and higher-order relations~\cite{Casarin2018,Lambiotte2019}, 
and temporal systems~\cite{Holme2012}. Due to the numerous applications of network science, the dynamic behavior of growing networks has absorbed much attention.

As an empirical example of why growth dynamics of a system are vital to investigate is the instability analysis and risk measurement in banking networks~\cite{Bardoscia2017}. Even with constant leverages overtime, when the number of involving agents in a financial network surges, the whole network may become unstable. Also, the increasing amount of leverage rates, linkages, and contracts among the agents can contribute to a higher instability~\cite{Bardoscia2017}. However, the implications of high contribution and connectedness throughout several scientific approaches have been discussed~\cite{Buld2019}.

We investigate the growth dynamics where a trade-off between the \textit{value} and the \textit{cost} emerges. Firstly, we consider a network in which its growth is affected based on the \textit{node activity}~\cite{Jafarietal}. The \textit{value} and the \textit{cost} are separately described by first-order coupled differential equations, then we obtain a modified preferential attachment of the BA model. Applying this modified model to such a network helps to clarify and predict the circumstances in which the network stops growing or continues to grow, or, there may exist some trade-off. Two types of drivers have been already considered~\cite{Jafarietal}. These drivers play a role in the network's dynamics such as node activity and also memory effects. The memory and the \textit{cost} play the role of friction against the growth of the network. They found a critical time-scale,--`characteristic time'-- when the attachment regime alters. Besides these, they found that the high event-wise temporal density --which implies higher cumulative degree, or similarly, the cumulative sum of past participation of a node in making links-- causes more distinct and distinguishable critical time.  As a whole, the size and the growth of networks are significant and influential to determine the future response of the systems.\\
In our study, we discuss 3 main concepts such as;\\
- the growth failure at the beginning;\\
- the permanent growth;\\
- the trade-off between stopping or continuing to grow.

From a microscopic point of view, on the contrary to investigation on the evolution of eigen constituents for extracting information of dynamic behaviors and trends reversal~\cite{Jurczyk2017},
we introduce differential equations for the \textit{value} and \textit{cost} of attachments. Due to the microscopic and the internal dynamics (\textit{value} and \textit{cost}) in a competitive atmosphere, each node interacts with the newcomer links. We evaluate the behavior of possible wandered links of the newcomer temporal nodes. This network is considered as a permanently growing situation, or no further growth situation, or living in a trade-off among no growth and permanent positive growth. This idea makes sense because in a non-collaborative economic world \cite{Grauwin2009} constituents seek to maximize their utility. Practically speaking, evaluating corporate strategies without considering the customers' earned \textit{value} and \textit{cost} aspects is not precise.
On the contrary to a discrete evaluation of just failure or just growing, we evaluate the trade-off between the \textit{value} and the \textit{cost} of the next attachments.
A question arises here. Why do we seek \textit{value} and \textit{cost} of creating links in a network? In the marketing strategies, when a firm launches a business platform, the managers may over time find it not \textit{cost} effective and they may leave it.
The managers may find that the \textit{cost} is more significant rather than the \textit{earned value} in terms of further growth. This consciousness has a great deal of importance.
As soon as the links get to know that the \textit{cost} of presence in the network is more than the \textit{value} which is earned, they will no longer be motivated for presence in the network and they are not likely to attach it. Once the \textit{cost} gets larger than the \textit{value}, the whole network stops growing. With the knowledge about \textit{value} and \textit{cost} behavior, one can describe possible scenarios in network's growth.

Based on the above-mentioned literature, one should be able to know the possible scenarios in the growth dynamics.
Some scholars have considered the \textit{value} and \textit{cost} as growth's drivers in the network. They assessed Nash equilibrium throughout the networks with different topologies. They designed a mechanism of information transmission with costly link creation. The value obtained by making a new link is information flow. Also, they considered different likelihoods of link elimination. This heterogeneous approach creates a wider range of equilibrium outcome~\cite{Sudipta2005}.

Other methods for growth dynamics are popular, namely, the effects of
connectivity based on the age~\cite{Krapivsky2000}, the nonlinear attachment probability~\cite{Krapivsky2001}, the evolution of competition in a network containing nodes with unique constant capability of absorbing new links~\cite{Bianconi2001}. Some popular methods for creating links include selecting the adequate connections to intensify collaboration among the weaker agents and also connections based on the networks' preferences not only the nodes' preferences, directed or undirected \cite{Iranzo2016}, but also the connections among the \textit{peripheral} and \textit{central} agents~\cite{Aguirre2013}.
Some other researchers studied the networks with accelerating growth and aging effects~\cite{Liu2019,Dorogovtsev2002,Ikeda2017,Safdari2016}, and with the evolutionary games under dynamic strategies in a finite population~\cite{Vardanyan2020,khalighi2019optimized}.

\section{Preferential Attachment}
The concept of preferential attachment is introduced\cite{Barabaacutesi1999,Barabsi1999,Newman2001} to state the tendency of nodes to make links with higher degree nodes. On the other side, the scholars~\cite{Dorogovtsev2002}
investigated the effects of sudden intentional random attack and damage (preferential elimination of nodes with the highest degree which means more exposure to be targeted). Hence, the network grows and is suddenly exposed to damage. This \textit{preferential damage} is against the \textit{preferential attachment}. The calculation of random damage threshold (which leads to the network's breakdown) was initially evaluated in~\cite{Cohen2000}. They analytically showed that by applying a preferential removal of a fraction of nodes, the percolation transition of breakdown will be dependent on the power-law and network size.
Another approach for investigating breakdown's predecessors is working on the phenomenon of removing links rather than removing influential nodes~\cite{Zhang2015}. A typical preferential attachment in a multi-graph system is shown to emerge under some circumstances in one network, and then it will lead to an increase or elimination of asymmetries in the whole system~\cite{Sudipta2020}.
To consider the deviations between theoretical aspects and real data, some scholars have considered a topological point of view such as the clustering coefficient, and have shown that there do exist some deviations among the real-world data and the theoretical networks. To solve this issue, the higher-order clustering coefficient is introduced~\cite{Fronczak2002}. Also, numerical analysis of real data turns up to highlight the memory effects and a critical time in cluster growth with a preference exponent related to memory strength. Hence, a preferential cluster growth exists in collective behaviors~\cite{Chmiel2013}.
\section{Trade-off Phenomenon}
Some scholars~\cite{Grauwin2009,Iranzo2016} investigated the networks containing sub-networks which include strong (high centrality) and weak (low centrality) networks in a game of achieving the strength (more centrality) and resulted that under circumstances based on Nash equilibrium and eigenvector centrality. In their model, the trade-off for being the strongest besides bearing the threats (such as a situation in which weaker competitors cooperate to overcome a certain stronger network) leads to a characteristic time. Hence, in the future of the competition, the dominant community is in danger of some disturbing dynamic costs such as \\
- originated internally (like costly structures or conflict of inner interests due to \textit{Agancy Theory} in some growth level of financial and social systems~\cite{Chari2019,Naeem2019};\\%
- or, originated externally (like competitors changing strategy and underestimating the collective profitability of the united weaker ones~\cite{Iranzo2016} in \textit{statue quo}).\\
A latent factor of these dynamics is agents' utility function which in some scales will finally form the future state of the networks~\cite{Grauwin2009}.
\section{Methodology}
BA preferential attachment states that nodes with more links are more exposed to be joined to the new links~\cite{Barabaacutesi1999,Barabsi1999}.
Initially, there exist $N$ nodes with initial \textit{Node Activity} of $k_0$. For creating an evolving process, at each time step, $m$ nodes among $N$ nodes are selected based on a uniform random distribution. Then, each of the selected nodes attaches to its destination node with the probability proportional to its node activity \cite{Jafarietal}. A certain node with higher node activity is more probable to attract a new attachment. We will have:
\begin{equation}\label{eq7}
k_i (t)= \sum_{t=0}^{t} L_i (t);
\end{equation}
where $L_i(t)$ means the number of links added to the node $i$ during $t$ to $t+dt$ time interval.
Considering the initial number of links in the network as $m_0$, after $t$ time-steps, we will have $m0+mt$ links.
Hence, during an `intermediary process' (adding links between constant number of previously existed nodes)\cite{Ikeda2017}, the evolving rate of node activity is as follows \cite{Jafarietal}:
\begin{equation}\label{eq8}
\frac{dk_i (t)}{dt}=m+\frac{m k_i (t)}{\sum_{j \neq i}^{N} k_j (t)}.
\end{equation}
Since each link possesses two ends, the attachment of one end is selected by a random uniform distribution. However, the attachment of the other end is created by the preferential attachment based on the higher activity~\cite{Jafarietal}, Eq.~\ref{eq7}. At each time step, the node with higher activity obtains a higher rate of attachment to it by the newcomers.
Consequently, there would be a competition among nodes for attracting more links. Eq.~\ref{eq7} and~\ref{eq8} show that during the contest, the nodes with higher activity in the network, have probably more chance to create links rather than the nodes with less activity.\\
By applying an analytical solution to above discussions, Eq.~\ref{eq11} is provided for value behavior:\\
\begin{equation}\label{eq11}
k_{v,i} (t)=2mt + c \sqrt{t};
\end{equation}	
where $c$ is constant and depends on the initial conditions as below:
\begin{equation}\label{eq12}
c=\frac{k_0 -2mt_0}{\sqrt{t_0}}
\end{equation}
In Eq.~\ref{eq11}, if ${t \to \infty}$ the network's behavior is more dominant by $2mt$. For ${t \to 0}$, $c\sqrt{t}$ will be more dominant.
\begin{equation}\label{eq13}
\begin{aligned}
\begin{cases}
&t \to \infty : k_v \sim t\\
&t \to 0	  : k_v \sim \sqrt{t}
\end{cases}
\end{aligned}
\end{equation}
There is a \textit{cross-over} time, ${t^*}$, which is simultaneous with the \textit{transformation} of value behavior~\cite{Jafarietal}. This cross-over time is a separating point between two behavioral regimes in Eq.~\ref{eq11} and it is calculated by Eq.~\ref{eq14}:
\begin{equation}\label{eq14}
t^{*}=\frac{c^2}{4m^2}.
\end{equation}
The behavioral transformation of \textit{value} indicates that by increasing $m$, while the network crosses its cross over, $t^*$, the growth of the network increases with a greater \textit{acceleration} \cite{Jafarietal,Dorogovtsev2002}. Since total number of links grows nonlinearly faster rather than linearly passing time steps~\cite{Ikeda2017}, this phenomenon is a ``cumulative growth''.
Noteworthy, each agent has its own \textit{value} and \textit{cost} equations (Eq.~\ref{eq0} and Eq.~\ref{eq17}).
Also, the fate of network will yield to the situations below:
\begin{equation}\label{eq0}
\begin{aligned}
\begin{cases}
k_v(t)-k_c(t)>0     : \textit{Growing (before $t^{characteristic}$);}\\
k_v(t)-k_c(t)\leq 0	: \textit{No Growth (after $t^{characteristic}$);}\\
\end{cases}
\end{aligned}
\end{equation}
where $k_v(t)$ and $k_c(t)$ refer to \textit{value} and \textit{cost} of making links, respectively.
We consider that the links not only grow due to \textit{value}, but also they are exposed to the \textit{costs} of attachments which also change by time. Our model considers changes of degrees in terms of the existence of \textit{cost}. $\alpha$ is the phase space. We will have~\cite{Jafarietal}:
\begin{equation}\label{eq17}
\begin{cases}
k_\nu (t) = 2mt + c \sqrt{t}\\
k_c (t)= (\alpha + m) t
\end{cases}
\end{equation}
\textit{Characteristic time} is a temporal moment which the \textit{cost} curve intersects the \textit{value} curve. At this moment, the \textit{value} and the \textit{cost} are equal, $k_v = k_c$. Hence, in the intersection we have:
\begin{equation}\label{eq18}
\begin{aligned}
&2mt+c\sqrt{t}= (\alpha +m)t \\
&t_{characteristic}=(\frac{c}{\alpha-m})^2.
\end{aligned}
\end{equation}

\section{Results}
\begin{figure}[htbp] 
	\centering
	\includegraphics[width=0.40\textwidth]{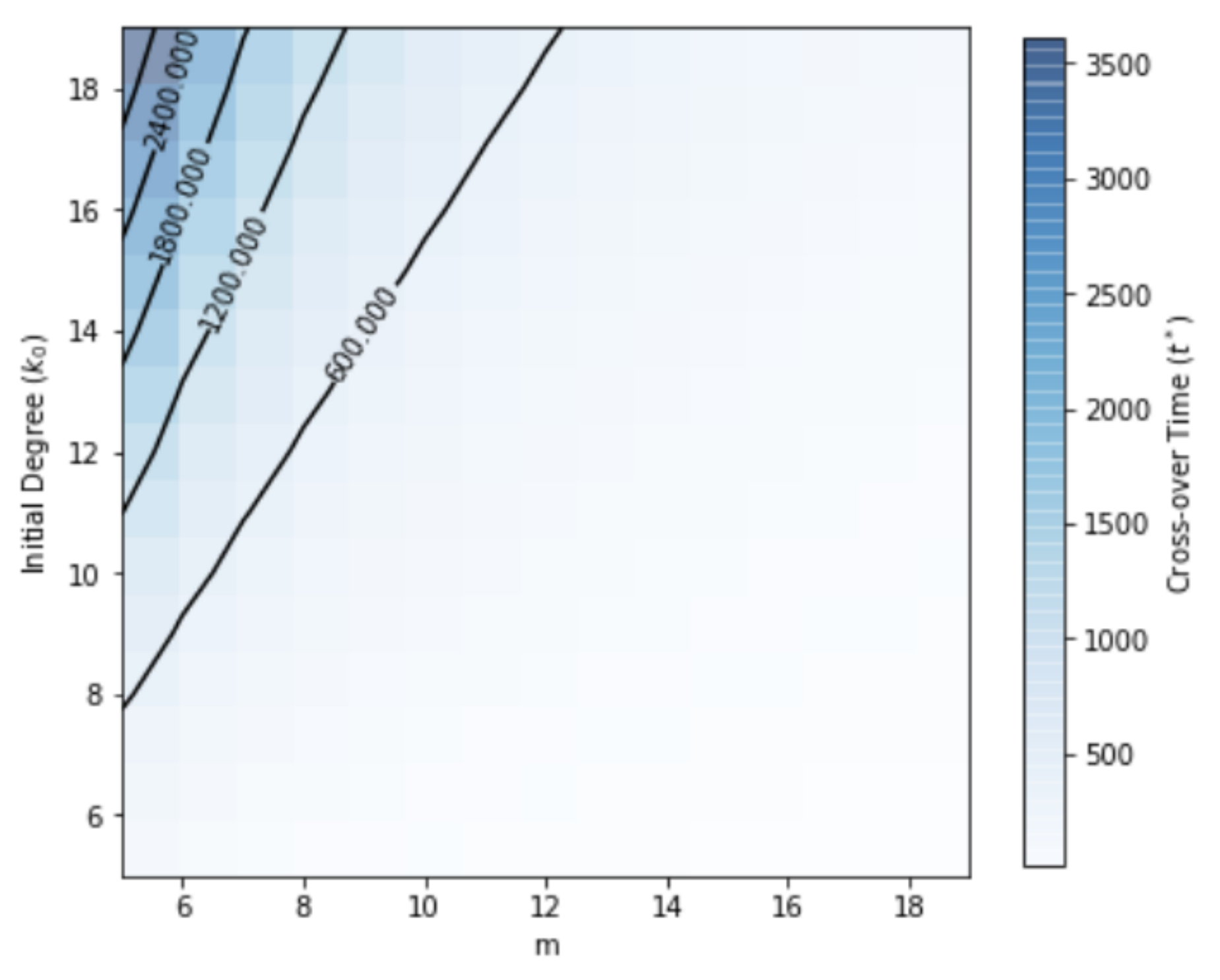}
	\caption{Cross-over time, ${t^*}$, for different initial degrees of the network, `$k_0$', versus the number of links, `$m$', which is created at each time step is demonstrated. Accordingly, the variations of $k_0$ and $m$ for $t^*=$ 600, 1200, 1800 and 2400 is depicted.}
	\label{fig_cross}
\end{figure}
\begin{figure*}[t] 
	\centering
	\includegraphics[width=0.7\textwidth]{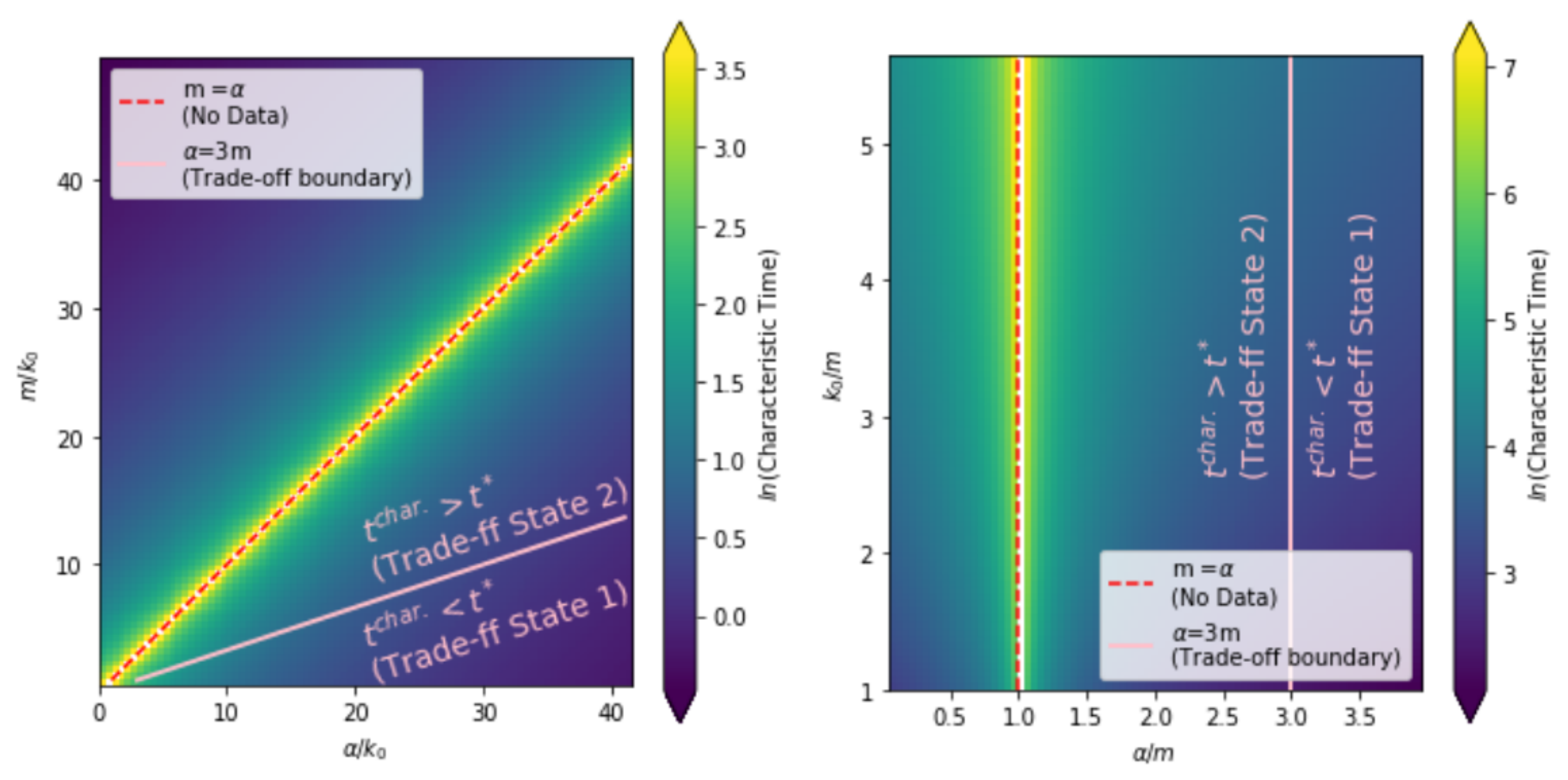}
	\caption{Left) Contour plot for $\ln$(Characteristic Time) considering the changes of $\frac{m}{k_0}$ versus the changes $\frac{\alpha}{k_0}$. Right) Contour plot for $\ln$(Characteristic Time) considering the changes of $\frac{k_0}{m}$ versus the changes $\frac{\alpha}{m}$.}
	\label{fig3}
\end{figure*}
\begin{figure}[tp] 
	\centering
	\includegraphics[width=0.5\textwidth]{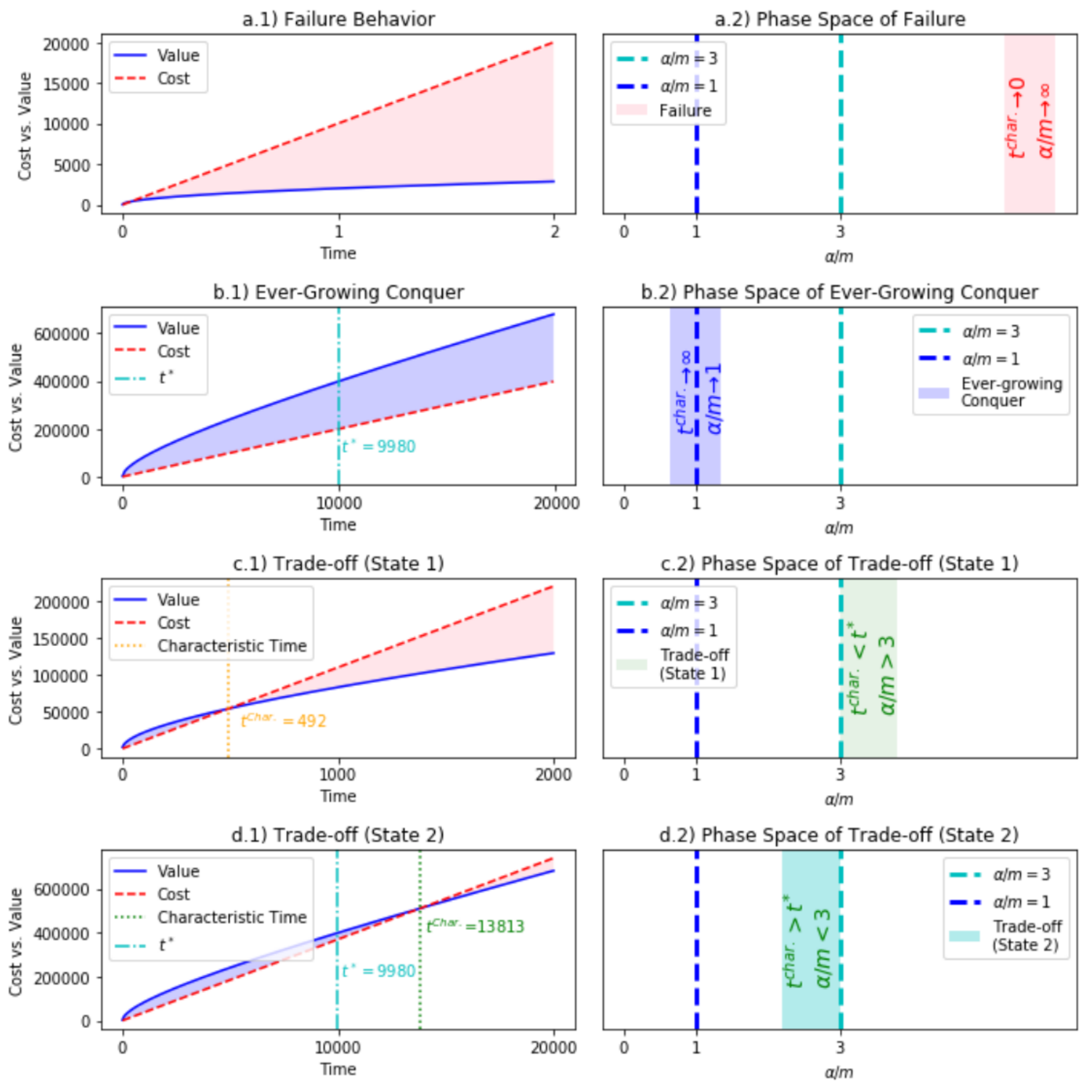}
	\caption{a) Scenario of Failure, b) Scenario of Ever-growing Conquer, c) Scenario of Trade-off in the 1st State, d) Scenario of Trade-off in the 2nd state.}
	\label{fig_scenario}
\end{figure}
Fig.\ref{fig_cross} represents \textit{cross-over} time, ${t^*}$, changing by different $k_0$ and $m$. When the initial degree $k_0$ is small, the increase in the number of links $m$ in each time step shortens ${t^*}$. On the other hand, the higher $k_0$ needs a higher link creation, $m$, to shorten $t^*$. This finding implies that the initial degree, $k_0$, acts like the inertia against the cumulative growth. Accordingly, it takes more time-steps for the network to pass the cross-over time of $t^*$.
The rise of $m$ causes the cumulative growth of the network to occur sooner. Lowering the cross-over time causes the network to escape from initial failure (Fig.\ref{fig3}).
In Fig.\ref{fig3}, the \textit{characteristic time} and its contours are illustrated for $m/k_0$ vs. $\alpha/k_0$ (left panel), and $k_0/m$ vs. $\alpha/m$ (right panel). Since lack of characteristic time is the best scenario for the growth, it is vital for the network to encounter the link creation factor $m$ equal to $\alpha$. When $k_0 \to 0$ (or $m/k_0 \to \infty$ and $\alpha/k_0 \to \infty$), the left and right panels together clarify the situation. As shown in the right panel of Fig.\ref{fig3}, when $k_0 \to 0$ (or $k_0/m \to 0$) the characteristic time will interestingly be more sensitive to the changes of $\alpha/m$ (around $\frac{\alpha}{m}=1$) rather than larger $k_0/m$. Hence, when the initial degree $k_0$ is extremely low, the trade-off between $\alpha$ and $m$ is more crucial. On the other hand, when the initial degree $k_0$ is extremely high, the aforementioned trade-off among $\alpha$ and $m$ is less sharper. It leads us to the consciousness that the higher initial degree at the network's birth lowers the sensitivity of network to future $\alpha$ parameter. As a typical rule for the trade-off, a high difference among $\alpha$ and $m$ lowers the characteristic time. Hence, the life of network will be shortened, as Fig.\ref{fig3} shows. Conversely, when $\alpha \to 0$, then $t^{Characteristic} \to \infty$. Fig.~\ref{fig3} shows that due to the node activity, a higher $m$ contributes to a shorter cross-over time, ${t^*}$. This implies that by increasing the ability of nodes to find each other, the evolution of network proceeds faster. Hence, the network will reach the cross-over time sooner.

When it comes to the real-world networks, faster information transmission and links creation will contribute to higher frequency in business cycles. In such markets, economic firms should lower their inertia internally and externally and permanently plan to develop new features on their goods and services. This will lower their network's \textit{cost} toward growth and postpone occurrence of \textit{characteristic time} in the state of trade-off between growth of \textit{value} and \textit{cost}.
According to the issues raised, one may consider 3 scenarios for the network's growth. Plus, the phase space relating to each scenario is presented in Fig.~\ref{fig_scenario}:\\
\textbf{Scenario of Failure:} In this scenario, the cost of making new links among temporal nodes is always more than the value. Hence, $t^{Characteristic} \to 0$. This phenomenon occurs for $\alpha>>m$, as panel \textit{a.2} in Fig.~\ref{fig_scenario} proves.\\
\textbf{Scenario of Ever-growing Conquer:} As shown in panel \textit{b.1} and \textit{b.2} of Fig.~\ref{fig_scenario}, the network has successfully fulfilled all the circumstances. In other words, the \textit{value} behavior has accelerated enough, and during `all' circumstances the \textit{cost} is less than the \textit{value} which is earned by making links. Accordingly, the constituents totally are satisfied! Hence, the network `continuously' grows. In the two areas along $\frac{\alpha}{m}$, the network has the chance of ``Ever-growing Conquer''. These areas conclude $\frac{\alpha}{m} \to 1$, as Eq.~\ref{eq18} and panel~\textit{b} of Fig.~\ref{fig_scenario} prove. \\
\textbf{Scenario of Trade-off:} The trade-off scenario can be explained by two events which both occur between failure and ever-growing scenarios. By equalizing the \textit{characteristic time} and the \textit{cross-over time}, $t^*$, we have two answers:
\begin{equation}\label{eq21}
\begin{cases}
\alpha = 3m\\
\alpha = -m
\end{cases},
\end{equation}
where $\alpha = -m$ is ineligible. Hence, two states emerge:\\
- Fig.~\ref{fig_scenario}, panel \textit{c},  shows that before acceleration of the \textit{value} behavior, the \textit{cost} outpaces the \textit{value} of making links. Hence, before that the network has the chance of changing the \textit{value} behavior, its growth stops.
The phase space is:
\begin{equation}\label{eq20}
t_{characteristic} < t^* \to \alpha > 3m.
\end{equation}
- Again, based on Fig.\ref{fig_scenario}, panel \textit{d}, the network may pass the value cross-over and consequently, it is able to rise acceleration of the \textit{value}. After passing the value's cross-over, the \textit{cost} of making newcomer temporal links boosts insofar as the \textit{cost} outpaces the \textit{value}. Eventually, the network's growth stops. This phenomenon may not occur soon enough to avoid the state of ever-growing. It actually depends on the targeted period. The phase space is:
\begin{equation}\label{eq19}
t_{characteristic} > t^* \to \alpha < 3m.
\end{equation}
In the stock market, this is the situation that the investors do not prefer to stay in the trading position. In the marketing, the business cycle is started to downturn and will be no longer a boosting industry as it was in the past, and the newer competitors may refuse to enter the industry. It is notable that the scaling features are significant in the trade-off scenario. To some scales, the network's growth can be a cumulative positive amplifier for a further growth, and on the other scales it may be as a barrier for a further growth. To avoid the network to be vulnerable to ``failure'', it can be kept safely in $0\le\frac{\alpha}{m}\le3$.

The existence of a friction-like factor (such as memory \cite{Jafarietal} and \textit{cost}) leads the system to be a smaller network. Intuitively, the emergence of wanderer agents, their memory, and the network's ability to absorb them, will determine the fate of the whole.
\section{Conclusion}
We found that some networks may fail at the beginning phase since their \textit{cost} overcomes their \textit{value} from the early beginning. Some are successful at the beginning phase, however, they stop growing before reaching the positive acceleration of value behavior (cross-over phenomenon). The trade-off scenario happens when the network successfully pass their beginning phase. However, they are exposed to failure after the emergence of \textit{characteristic time}, because their network's \textit{cost} of growth overcomes its \textit{value} simultaneous to the characteristic time. The nodes' interest may be latent and be against collective motivations~\cite{Grauwin2009} of sub-networks and communities.

Our proposed model implies that the coexistence of cumulative \textit{value} in preferential attachment versus \textit{cost} of attachments may cause the network to be not ever-growing. Then, under some circumstances, the growth stops. Hence, depending on the quality of changes, a trade-off scenario is possible. In some scales, the network's growth accelerates itself and in others, it may cause some barriers for further growth.
The increase in the number of links around certain nodes may cause the emergence of monopoly~\cite{shirazi2013}. On the other hand, the network may be more vulnerable to the \textit{cost} of any further growth. Thereby, the competitors may be more focused on the network. The attraction of competitors can be considered as the attachment \textit{cost}. In the long term, the attraction of the proposed network decreases. Hence, the probable links of newer time-steps may diverge from preferential attachment, and the \textit{cost} effects become more significant. Our proposed model can be employed on the systems with finite growth and multi-criteria drivers such as financial and marketing industries.

\bibliographystyle{elsarticle-num}
\bibliography{NetworkGrowth.bib}

\end{document}